\documentclass[letterpaper, 10 pt, conference]{ieeeconf}  

\IEEEoverridecommandlockouts                              
\usepackage{amsmath} 
\usepackage{amssymb}  
\usepackage{multirow}
\usepackage{xcolor}
\usepackage{graphicx}
\usepackage{subcaption}

\newcommand{\pa}{p}
\newcommand{\ua}{u_A}
\newcommand{\ud}{u_D}
\newcommand{\sigmaa}{\sigma_A}
\newcommand{\sigmad}{\sigma_D}
\newcommand{\type}{\theta}

\newcommand{\Ex}{\mathbb{E}}
\DeclareMathOperator*{\argmin}{\arg\!\min}
\DeclareMathOperator*{\argmax}{\arg\!\max}

\newtheorem{lemma}{Lemma}

\renewcommand{\QED}{\hfill\QEDopen}

\title{\LARGE \bf Strategic Inference with a Single Private Sample}

\author{Erik~Miehling, Roy~Dong, C{\' e}dric~Langbort, and Tamer~Ba{\c s}ar
\thanks{This work was supported by the U.S. Office of Naval Research (ONR) MURI grant N00014-16-1-2710.}
\thanks{All authors are with the Coordinated Science Lab, University of Illinois at Urbana--Champaign, Urbana, IL 61801, USA.
        {\tt\footnotesize \{miehling,roydong,langbort,basar1\}@illinois.edu}}%
}

\begin{document}

\maketitle
\thispagestyle{empty}
\pagestyle{empty}

\begin{abstract}
Motivated by applications in cyber security, we develop a simple game model for describing how a learning agent's private information influences an observing agent's inference process. The model describes a situation in which one of the agents (\emph{attacker}) is deciding which of two targets to attack, one with a known reward and another with uncertain reward. The attacker receives a single private sample from the uncertain target's distribution and updates its belief of the target quality. The other agent (\emph{defender}) knows the true rewards, but does not see the sample that the attacker has received. This leads to agents possessing asymmetric information: the attacker is uncertain over the parameter of the distribution, whereas the defender is uncertain about the observed sample. After the attacker updates its belief, both the attacker and the defender play a simultaneous move game based on their respective beliefs. We offer a characterization of the pure strategy equilibria of the game and explain how the players' decisions are influenced by their prior knowledge and the payoffs/costs.
\end{abstract}

\section{INTRODUCTION}

As our societal systems become increasingly populated with smart devices, there is an increasing number of learning agents collecting data and drawing inferences. Furthermore, as more and more decisions are being made by (automated) learning agents, an increasing number of secondary agents will arise to monitor their decision making processes.

The associated problem that one faces when learning about the relevant environment becomes much more complex in such settings. Agents no longer learn about their environment in isolation; an agent's learning process (the act of collecting information) is subject to observation from others who are attempting to infer their private information. 
Since beliefs influence actions, and actions influence payoffs, each agent must be mindful of how its learning actions influence what others believe about its private information. This is especially a concern when the interests of the agents are misaligned.\footnote{The influence of one's actions/strategies on the beliefs of others is known in the economics literature as \emph{signaling}. Signaling is present in both competitive (game) settings \cite{tavafoghi2018unified2} and cooperative (decentralized control) settings \cite{ho1978teams,tavafoghi2018unified1}.} As a result, a complete analysis of how an agent should learn in such environments must explicitly take into account how the learning process itself influences players' beliefs and their subsequent choices.

The general problem of learning while under observation arises in many real-world scenarios. Examples range from problems in e-commerce and online marketplaces, \emph{e.g.}, where a merchant estimates a user's willingness to pay based on their browsing patterns in order to set revenue-maximizing prices, to cyber security, \emph{e.g.}, where a defender partially observes an attacker's reconnaissance of various target locations in order to determine where to allocate defensive resources.

For the purposes of this paper, our focus is on cyber security. As a motivating scenario, consider a hacker carrying out some preliminary reconnaissance on potential attack vectors, \emph{e.g.}, sending out commands to a particular server to see which ports are open, before launching an attack. Such recon actions not only provide information to the hacker about the structure of the network but also, due to monitoring systems deployed throughout the system, reveal information to the defender about the hacker's knowledge and true intent. The defender thus faces a problem of attempting to infer the hacker's state of knowledge based on the recon actions. This is further complicated by the fact that the defender may only partially observe the recon actions, for instance, while the defender can determine which server received a malicious command, it may not know what information the hacker had received. Consequently, the defender does not know the hacker's updated state of knowledge. Conversely, the hacker would be unsure of the value of various attack vectors/targets, and thus does not know how its private information influenced the defender's view. The interaction is thus one of asymmetric information: each agent possesses information not available to the other. For the remainder of the paper, we will present the results in the context of cyber security and refer to the agents as the attacker and the defender.

A first step in analyzing this problem is understanding how the learning agent's private information influences the observing agent's inference process. In this paper, we introduce a stylized game-theoretic model to capture the basic strategic components of this interaction in the context of the previously described cyber security scenario. Specifically, we consider an attacker and a defender who simultaneously choose among two targets, $a$ and $b$. If the attacker chooses to attack $a$, it receives a randomized reward with an uncertain distribution; if the attacker chooses to attack $b$, it receives a fixed and known reward. In both cases, the defender loses the same amount. If the defender chooses to defend the same target that the attacker attacked, then the attacker incurs a cost of capture and the defender gains the same amount (a reward for the capture). The interesting part of this model is what happens beforehand. Before the attacker and the defender play the game, the attacker receives a single private sample from the randomized reward distribution. While the attacker's prior before this observation is common knowledge, the posterior after the observation is known only to the attacker. The defender knows the true distribution of the reward, but does not see the attacker's sample and thus does not know the posterior of the attacker. 
The fundamental problem we investigate is the influence of the attacker's private sample on beliefs and equilibria of the game.


\subsection{Background \& Related Work}


The structure of our model bears similarity to some existing models in the economics and learning literatures. 
One such model, termed \emph{strategic experimentation} \cite{bolton1999strategic}, describes a setting where agents learn from the outcomes of other's experiments. In the original model \cite{bolton1999strategic}, each experiment, defined as the selection of an alternative, yields an outcome that is commonly observed by the agents. The model is a multi-agent extension of the two-armed bandit\footnote{See \cite{mahajan2008multi} for an overview of multi-armed bandits and some fundamental results.} problem \cite{woodroofe1979one,berry1980two} where agents choose between a ``risky'' alternative (with unknown statistics) and a ``known'' alternative, with the objective of maximizing expected payoff. 
The main insight from the classical model is that, due to free-riding, any equilibrium in which players only use the common belief is socially suboptimal. 

The original strategic experimentation model has been extended in many directions. The most relevant to our setting are the models of \cite{rosenberg2007social,heidhues2015strategic} that consider the case of private information. Under private information, the outcomes of experiments are no longer publicly observed, rather they are assumed to be private to the players. 
The model of \cite{rosenberg2007social} studies a two player problem where each player faces a bandit problem consisting of a risky arm (of two possible types) and a known arm. Similarly, \cite{heidhues2015strategic} study a related problem but allow for communication (via cheap talk) among the players and for the players to reverse their decisions to stop experimenting, distinguishing it from the setup of \cite{rosenberg2007social}.

A related setting is the \emph{Bayesian exploration} model of \cite{mansour2016bayes}. The model considers a problem where a principal is attempting to incentivize (a series of) myopic agents to explore, rather than be greedy and exploit. This is similar in spirit to the motivation found in strategic experimentation settings.




Given the motivation for our study, it is useful to also discuss some models from the security literature. The general problem of allocating defenses in the presence of a strategic attacker is a well-studied topic. One prominent model is the \emph{Colonel Blotto game} \cite{borel1953theory,ferdowsi2018generalized,gupta2014three} consisting of two players who aim to allocate resources (troops) to alternatives (battlefields) in order to maximize wins (by majority rule). 
Another popular model is that of \emph{Stackelberg security games} \cite{sinha2018stackelberg} where a defender (leader) allocates defenses to a collection of targets before an attacker (follower) launches its attack. While differing in the timing and informational assumptions, both the Colonel Blotto and Stackelberg security game settings provide insight into how defensive resources should be allocated in order to minimize the chance of a successful attack.

Compared to the aforementioned work, the model presented in this paper studies a simple setting in which the learning agent receives a single sample of private information from a distribution privately known by another (observing) agent. Our model aims to isolate the effect of this single private sample on the inference process of the observing agent. This structured information asymmetry has some natural applications (some of which were discussed earlier), especially in the context of cyber security. To the best of our knowledge, models possessing this structure have not yet been investigated in the economics, learning, or security literatures.

Comparing our model to the more general strategic experimentation setting, our model is one of private information; however, we note that the learner is not actively deciding to receive the sample, it receives the sample regardless of its choice. As a result, we refer to the model of this paper as \emph{strategic inference} rather than strategic experimentation.

\subsection{Contribution}

In summary, the contribution of our work is a game formulation, motivated by the aforementioned foundational security setting, that isolates a structured class of asymmetric information games that admits tractable analysis. The model provides insight into how an observing agent reasons about the private information revealed to a learning agent and its subsequent impact on decisions.

\section{THE GAME MODEL}
\label{sec:gamemodel}

Consider the following two-player game where an attacker (A) and defender  (D) choose among two potential targets $a$ and $b$. The reward of target $a$ is uncertain to the attacker, dictated by an unknown distribution, whereas target $b$ has a known reward. The attacker is assumed to possess a prior over the unknown distribution. For the purposes of this paper, we assume a Bernoulli reward distribution (for the unknown target $a$) and a beta prior.\footnote{This choice is for mathematical simplicity (conjugate priors). The model can also consider alternative distributions, such as Gaussian.} The Bernoulli parameter, denoted by $\pa$, is only known to the defender, whereas the beta prior, denoted by $(\alpha_0,\beta_0)$ is assumed to be common knowledge. 

\begin{figure}[htbp]
\vspace{0.25em}
\begin{center}
        \includegraphics[width=0.85\columnwidth]{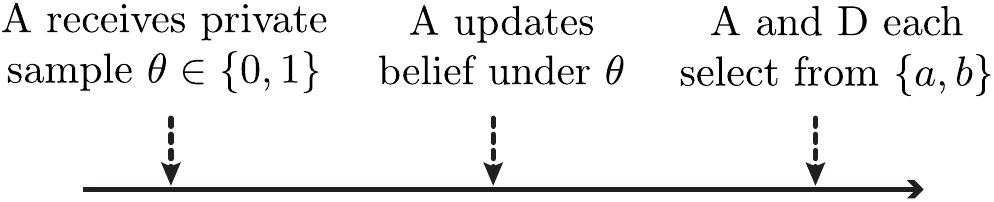}
\caption{Order of events in the game.}
\label{fig:timeline}
\end{center}
\vspace{-0.75em}
\end{figure}

The game proceeds as follows. First, the attacker receives a sample from the unknown target $a$, which it uses to form an updated (posterior) belief. Since the prior is assumed to be beta and the trials Bernoulli, the updated posterior is simply $\text{Beta}(\alpha_0+\type,\beta_0+1-\type)$. The defender does not see the realized sample $\type$. Players then act simultaneously, the attacker choosing which target to attack and the defender choosing which target to defend. Fig. \ref{fig:timeline} represents the timeline of this interaction.

The objectives of the attacker and the defender are misaligned; the attacker wishes to avoid the defender whereas the defender wishes to catch the attacker. 
These preferences are encoded using payoffs as follows. If the attacker selects target $a$, it receives a payoff of $R>0$ with probability $\pa$. If the attacker picks $b$, it receives a known payoff of $\gamma R$ where $0<\gamma<1$ is a constant. 
The attacker incurs a capture cost of $c>0$ if the defender picks the same target. Payoffs are zero-sum where the attacker is assumed to be maximizing.

\subsection{Subjective Payoffs \& Costs}
\label{ssec:subj}

Before describing payoffs, we define the players' types. The attacker's type is the sample it receives from target $a$, denoted by $\type \in \{0,1\}$. The defender's type is the true (reward) parameter of the reward distribution of target $a$, denoted by $\pa \in [0,1]$. Note that while the defender has private knowledge of the true $\pa$, it does not know the sample that the attacker received. In other words, neither player knows the other player's true type.

The attacker's uncertainty of the defender's type leads to it being uncertain of the true payoffs of the game. The attacker can only compute its expected payoff using its prior and the information it gathers from the received sample. Specifically, the attacker's best estimate of the true parameter $\pa$ is given by the mean of its posterior after the sample is received, thus its expected reward for selecting target $a$ is $\frac{\alpha_0+\type}{\alpha_0+\beta_0+1}R$. The defender on the other hand, knows the true expected costs, since it knows the reward parameter $\pa$. Furthermore, since $\gamma$ is known, the payoff for selecting target $b$ is known to both players. The players' subjective payoffs/costs are illustrated below in Table \ref{tab:sub}.

\begin{table}[h]
\caption{\textsc{Subjective payoffs/costs for players.}}

\vspace{-0.5em}
\begin{subtable}{\columnwidth}
\centering
\normalsize
\bgroup
\def\arraystretch{2.25}
\hspace{-4em}\begin{tabular}{llcc}
 &  & \multicolumn{2}{c}{\hspace{0.75em}$D$}\\ [-1.75em]
 &  & $a$ & $b$ \\ \cline{3-4} 
\multirow{2}{*}{$A$} & \multicolumn{1}{l|}{$a$} & \multicolumn{1}{c|}{$\frac{\alpha_0+\type}{\alpha_0+\beta_0+1}R-c$} & \multicolumn{1}{c|}{$\frac{\alpha_0+\type}{\alpha_0+\beta_0+1}R$} \\ [0.25em]\cline{3-4} 
 & \multicolumn{1}{l|}{$b$} & \multicolumn{1}{c|}{$\gamma R$} & \multicolumn{1}{c|}{$\gamma R-c$} \\ [0.25em]\cline{3-4} 
\end{tabular}
\egroup
\vspace{1em}
\caption{Subjective payoffs for the attacker, denoted $\ua$.}
\end{subtable}

\begin{subtable}{\columnwidth}
\centering
\normalsize
\bgroup
\def\arraystretch{1.8}
\hspace{-4em}\begin{tabular}{llcc}
 &  & \multicolumn{2}{c}{$D$}\\ [-1.25em]
 &  & $a$ & $b$ \\ \cline{3-4} 
\multirow{2}{*}{$A$} & \multicolumn{1}{l|}{$a$} & \multicolumn{1}{c|}{$\pa R-c$} & \multicolumn{1}{c|}{$\pa R$} \\ \cline{3-4} 
 & \multicolumn{1}{l|}{$b$} & \multicolumn{1}{c|}{$\gamma R$} & \multicolumn{1}{c|}{$\gamma R-c$} \\ \cline{3-4} 
\end{tabular}
\egroup
\vspace{1em}
\caption{Costs for the defender, denoted $\ud$.}
\end{subtable}
\label{tab:sub}
\vspace{-1em}
\end{table}

\subsection{Best Response Functions}
\label{ssec:BR}

The strategies of the attacker and the defender, denoted by $\sigmaa(a;\type)$ and $\sigmad(a;\type)$, respectively, represent the probabilities that target $a$ will be chosen. The best response functions of the players are given by the following optimizations.
\begin{align*}
	\sigmaa^*(a;\type)&\in\argmax_{\sigmaa\in[0,1]}\Big\{\Ex_{\pa}\big[\ua(\sigmaa,\sigmad^*(a;\pa))\mid\type\big]\Big\}\\
	\sigmad^*(a;\pa)&\in\argmin_{\sigmad\in[0,1]}\Big\{\Ex_{\type}\big[\ud(\sigmaa^*(a;\type),\sigmad)\mid \pa\big]\Big\}
\end{align*}
Note that the attacker is maximizing and the defender is minimizing, reflecting the fact that the attacker wants to avoid the defender but the defender wants to catch the attacker. Substitution of the subjective payoffs/costs from Table \ref{tab:sub} yields the following best response functions. 

\vspace{0.5em}
\begin{lemma}\label{lem:eqconds}
The best response functions of the players are
\begin{align}
	\nonumber\sigmaa^*(a;\type)&\in\argmax_{\sigmaa\in[0,1]}\Bigg\{\sigmaa\bigg(\bigg(\frac{\alpha_0+\type}{\alpha_0+\beta_0+1} - \gamma \bigg)R\\
	&\hspace{5em}+c\big(1-2\Ex_{\pa}\big[\sigmad^*(a;\pa)\mid \type\big]\big)\bigg)\Bigg\},\label{eq:strat1}\\
	\sigmad^*(a;\pa)&\in\argmin_{\sigmad\in[0,1]}\Big\{\sigmad\big(1-2\Ex_{\type}\big[\sigmaa^*(a;\type)\mid \pa\big]\big)\Big\}\label{eq:strat2}
\end{align}
where $\Ex_{\pa}\big[\sigmad^*(a;\pa)\mid \type\big]$ is the attacker's expectation of the defender's strategy and $\Ex_{\type}\big[\sigmaa^*(a;\type)\mid \pa\big]$ is the defender's expectation of the attacker's strategy.
\end{lemma}

\proof{ See Appendix \ref{app:lem:eqconds}. }

\vspace{0.25em}


The form of the expectations of the strategies in Lemma \ref{lem:eqconds} reflects the information asymmetry among the players: the attacker sees $\type$ but does not know $p$; the defender knows $p$ but does not see $\type$. The attacker computes the expectation of the defender's strategy, $\Ex_{\pa}\big[\sigmad^*(a;\pa)\mid \type\big]$, using its private sample. Given a beta prior, $\text{Beta}(\alpha_0,\beta_0)$, the attacker's distribution over the true reward parameter as a function of the private sample is dictated by a beta distribution, $\text{Beta}(\alpha_0+\type,\beta_0+1-\type).$ The defender computes the expectation of the attacker's strategy, $\Ex_{\type}\big[\sigmaa^*(a;\type)\mid \pa\big]$, given its knowledge of the reward parameter $p$. The defender's distribution over the attacker's private sample is given by a Bernoulli distribution, $\text{Bernoulli}(p)$. Since the attacker's type (sample) is generated from a probability distribution dictated by the defender's type (parameter of the reward distribution), the game is one with correlated types \cite{fudenberg1991game}. This structure is core to our analysis; under correlation, knowledge of one's type is informative for inferring the other player's type.

\section{Pure Strategy Equilibria}
\label{sec:equil}

As mentioned in the introduction, we are interested in studying pure strategy equilibria. The following lemma shows that, for a given set of parameters, $R$, $c$, $\gamma$, $(\alpha_0,\beta_0)$, the resulting game has a (unique) pure strategy equilibrium which can take three possible forms. 
\begin{lemma}\label{lem:equil}
The private sampling game has at most one pure strategy saddle-point equilibrium, characterized by the following three disjoint regions:
\begin{enumerate}
\item $\sigmaa(a;\type) =\sigmad^*(a;\pa) = 1$ for all $\type,\pa$ if and only if
\begin{align*}
	\frac{\alpha_0+\type}{\alpha_0+\beta_0+1}R -c \ge \gamma R
\end{align*}
\item $\sigmaa(a;\type) =\sigmad^*(a;\pa) = 0$ for all $\type,\pa$ if and only if
\begin{align*}
	\gamma R - c \ge \frac{\alpha_0+\type}{\alpha_0+\beta_0+1}R
\end{align*}
\item $\sigmaa^*(a;0) = 0$, $\sigmaa^*(a;1) = 1$, and\newline $\sigmad^*(a;p)= \mathbb{I}\{p>1/2\}$ if and only if
\begin{align*}
	I_{1/2}(\alpha_0,\beta_0) \in\\
	&\hspace{-6em}\Bigg[\frac{1}{2}\bigg(1+\frac{R}{c}\Big(\gamma-\frac{\alpha_0+1}{\alpha_0+\beta_0+1}\Big)\bigg) + \frac{\Big(\frac{1}{2}\Big)^{\alpha_0+\beta_0}}{\alpha_0B(\alpha_0,\beta_0)},\\ 
	&\hspace{-5.4em}\frac{1}{2}\bigg(1+\frac{R}{c}\Big(\gamma-\frac{\alpha_0}{\alpha_0+\beta_0 + 1} \Big)\bigg) - \frac{\Big(\frac{1}{2}\Big)^{\alpha_0+\beta_0}}{\beta_0B(\alpha_0,\beta_0)}\Bigg]
\end{align*}
where $B(\alpha,\beta)$ and $I_{x}(\alpha,\beta)$ are the beta function and regularized incomplete beta function, respectively. 
\end{enumerate}

\end{lemma}

\noindent\proof{See Appendix \ref{app:lem:equil}}.

\subsection{Equilibrium Discussion}

Our discussion proceeds in two steps. First, for fixed parameters $R$, $c$, and $\gamma$, we describe the $(\alpha_0,\beta_0)$ region where each equilibrium condition of Lemma \ref{lem:equil} holds. Second, we vary the capture cost $c$ to investigate how the equilibrium-supporting regions change.


We study the parameter regime in which $R-c > \gamma R$. The rationale for this choice is as follows. The quantity $R-c$ is the highest possible reward the attacker can receive from the uncertain target if it gets caught. However, the attacker does not always receive $R$ from $a$, it may get unlucky and get a zero reward (since the reward from $a$ is stochastic) in addition to getting caught. Due to the attacker's prior on the true value of $\pa$, the attacker's view of the reward from playing $a$ (when the defender also plays $a$), regardless of the received sample, is strictly less than $R-c$. By analyzing the equilibria in the $R-c > \gamma R$ regime, we can investigate the specific level of uncertainty (quantified by the prior) such that the attacker would prefer to play $b$ over $a$. As such, we fix $R=1$, $c=0.1$, and $\gamma = 1/2$ for the following analysis.

\begin{figure}[t]
    \centering
    \begin{subfigure}{0.375\columnwidth}
    \centering
        \includegraphics[width=\linewidth]{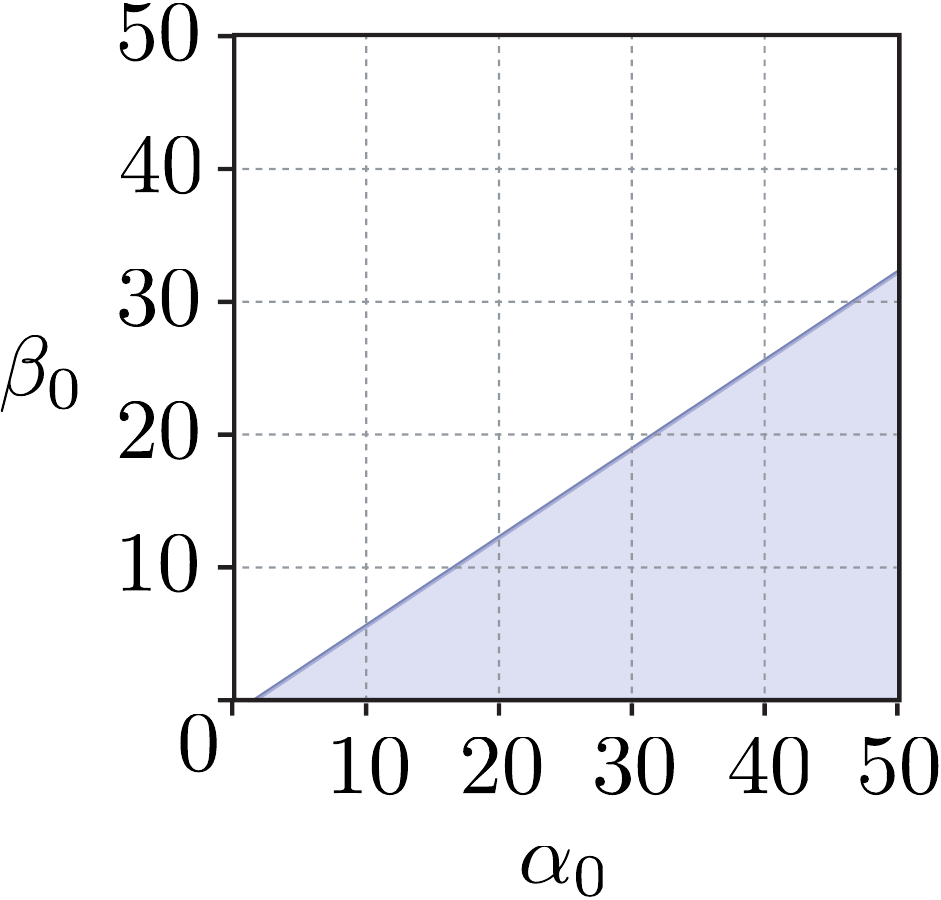}
        \caption{Region where both players choose target $a$.}
        \label{fig:ps12_1}
    \end{subfigure}%
    \hspace{0.15\columnwidth}%
    \begin{subfigure}{0.375\columnwidth}
    \centering
        \includegraphics[width=\linewidth]{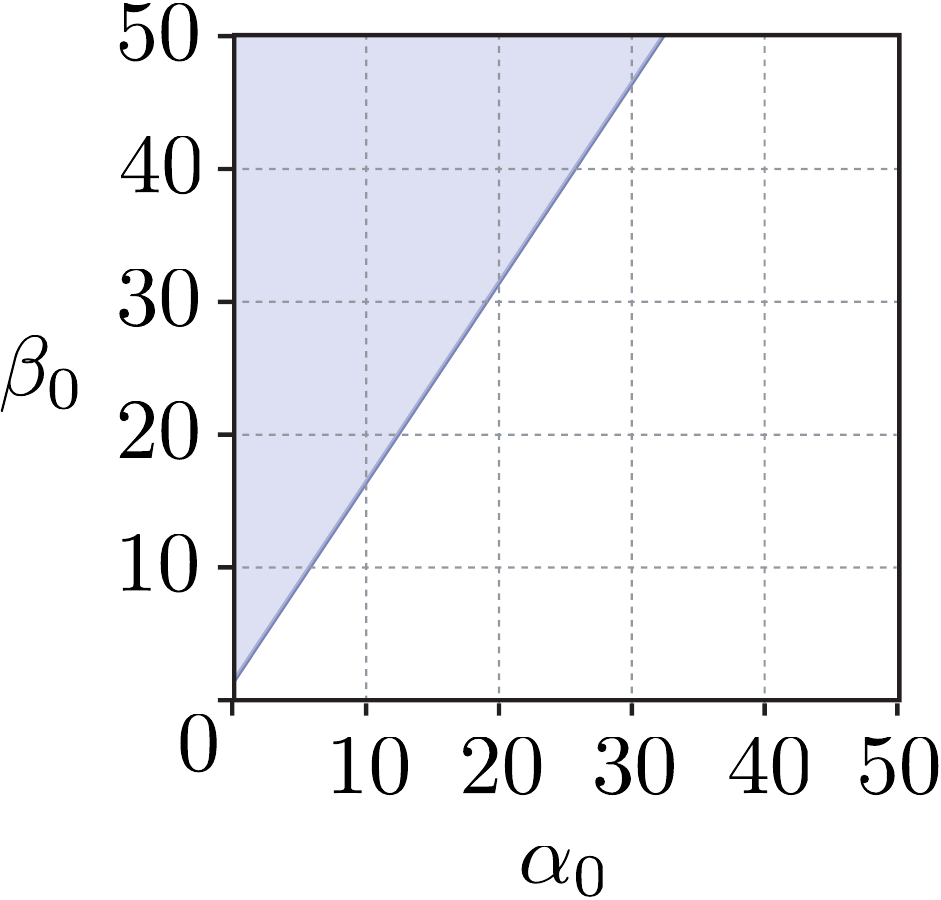}
        \caption{Region where both players choose target $b$.}
        \label{fig:ps12_2}
    \end{subfigure}
    \caption{Regions (blue/shaded) for $(\alpha_0,\beta_0)$ such that (a) both players choose target $a$; (b) both players choose target $b$.}
    \label{fig:ps12}
\end{figure}

For fixed $R$, $c$, and $\gamma$, the equilibrium conditions of Lemma \ref{lem:equil} describe regions in the space of prior parameters $(\alpha_0,\beta_0)$. In particular, the first two regions in Lemma \ref{lem:equil}, namely 1) $\sigmaa(a;\type) =\sigmad^*(a;\pa) = 1$ for all $\type$, $\pa$, and 2) $\sigmaa(a;\type) =\sigmad^*(a;\pa) = 0$ for all $\type$, $\pa$ are illustrated in Figure \ref{fig:ps12}. The interpretation of the first two regions is straightforward. Recall from Lemma \ref{lem:equil} the condition for the first region $\sigmaa(a;\type) =\sigmad^*(a;\pa) = 1$ is $\frac{\alpha_0+\type}{\alpha_0+\beta_0+1}R -c \ge \gamma R$. The equilibrium condition states that this must hold true for all received samples. Thus, regardless of the sample that the attacker receives, it must believe with a sufficiently high confidence that target $a$ will yield reward $R$. This is only true when $\alpha_0$ is much larger than $\beta_0$, characterized by the shaded region in Fig. \ref{fig:ps12_1}. The reasoning for the second region follows similarly (see Fig. \ref{fig:ps12_2}).

In the third region, the attacker follows its observation and the defender defends the more valuable target. That is, if the attacker receives a positive sample, $\type=1$, it attacks target $a$, and vice versa. The defender defends $a$ if and only if $\pa>1/2$. This equilibrium is best interpreted by separating the interval condition into its two inequalities. From the proof of Lemma \ref{lem:equil}, the condition for $\sigmaa^*(a;\type=1) = 1$ (the lower bound in the statement of the lemma) is
\begin{align*}
	I_{1/2}(\alpha_0+1,\beta_0) \ge \frac{1}{2}\bigg(1+\frac{R}{c}\Big(\gamma-\frac{\alpha_0+1}{\alpha_0+\beta_0+1}\Big)\bigg).
\end{align*}

\begin{figure}[h]
\begin{minipage}{0.245\columnwidth}
\begin{subfigure}{\linewidth}
\hspace{-0.5em}\includegraphics[width=\linewidth]{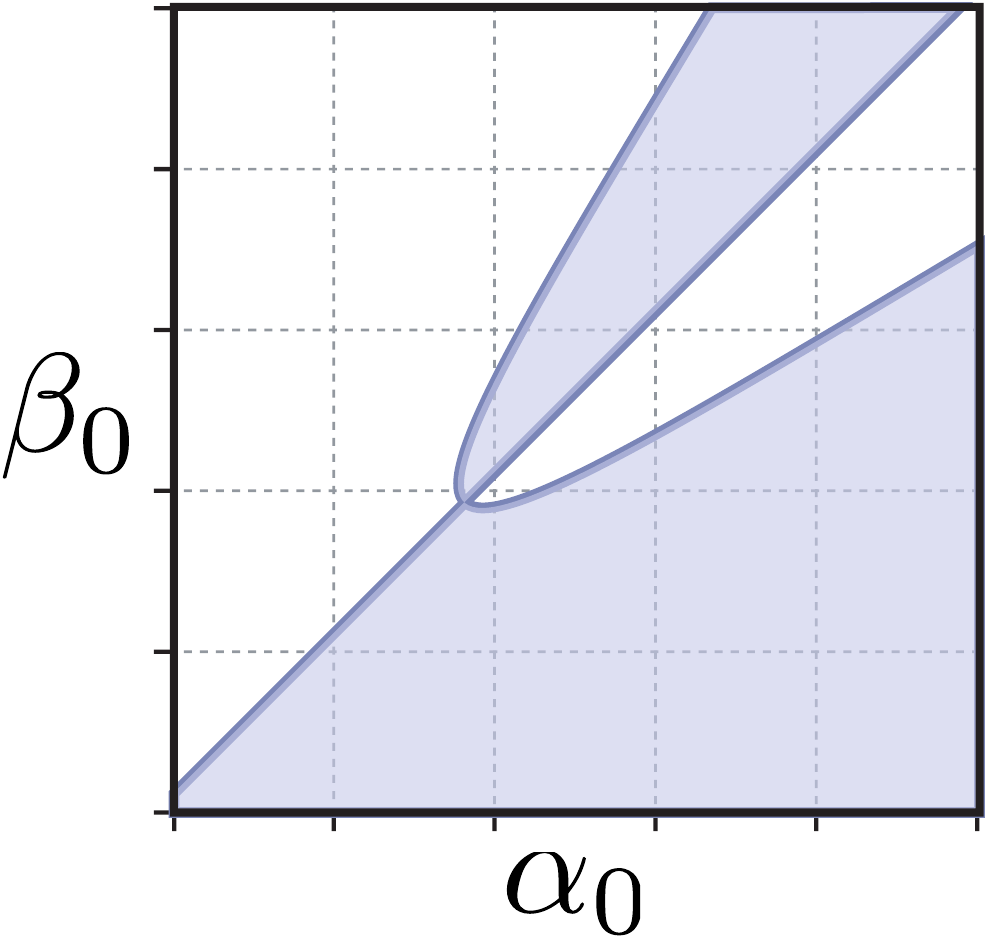}
\caption{}
\end{subfigure}
\begin{subfigure}{\linewidth}
\vspace{1em}
\hspace{-0.5em}\includegraphics[width=\linewidth]{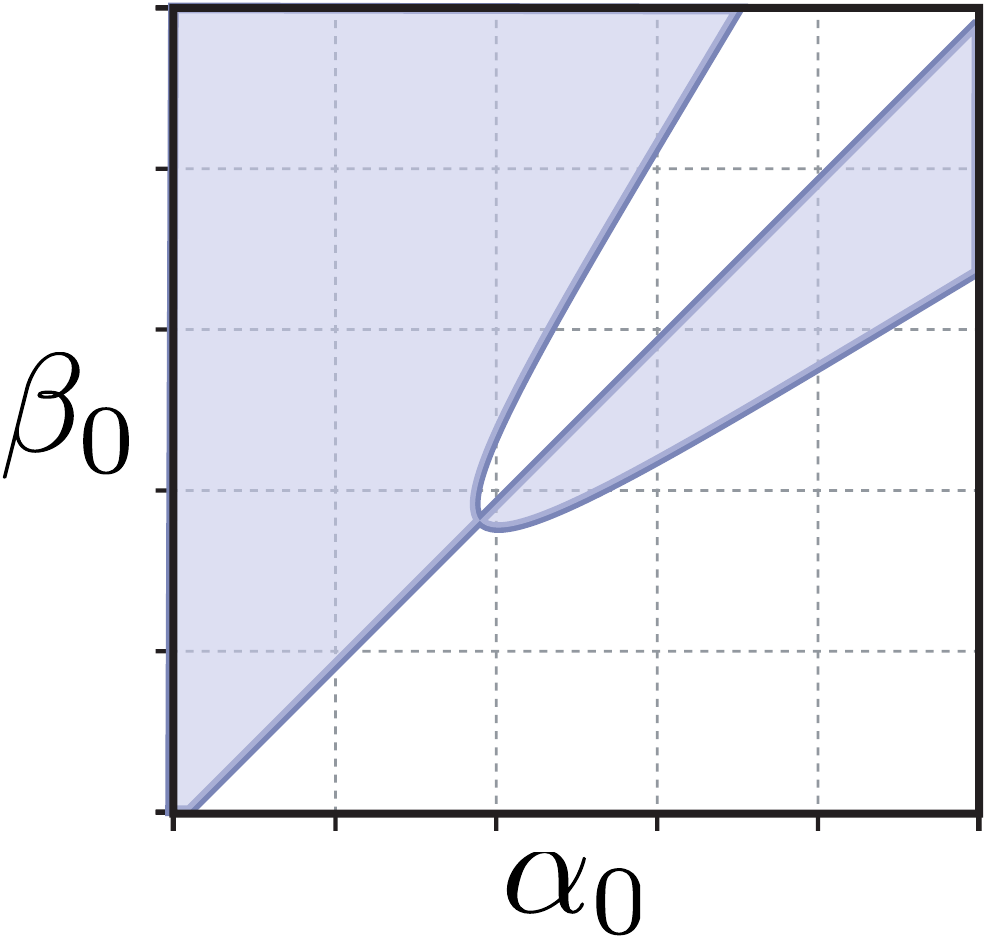}
\caption{}
\end{subfigure}
\end{minipage}%
\hspace{0\columnwidth}
\begin{minipage}{0.66\columnwidth}
\begin{subfigure}{\linewidth}
\includegraphics[width=\linewidth]{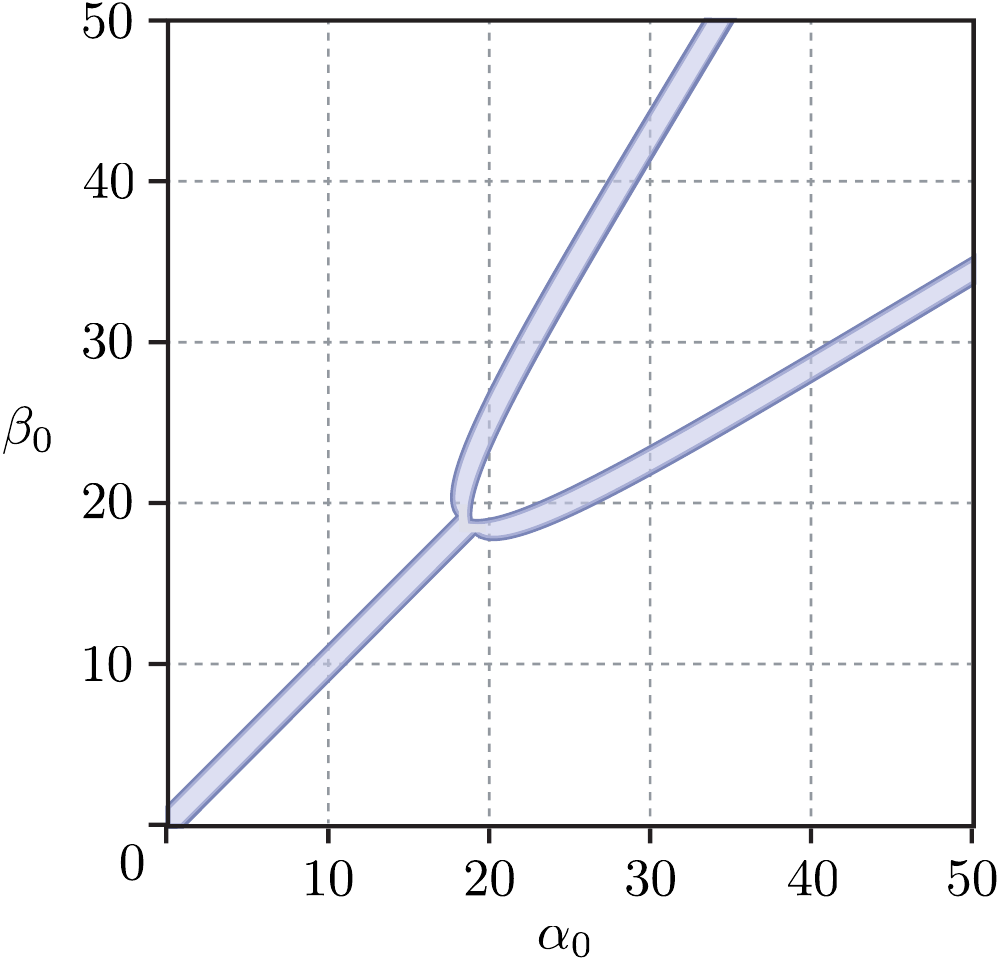}
\caption{}
\end{subfigure}
\end{minipage}
\caption{Region (shaded/blue area in (c)) for $(\alpha_0,\beta_0)$ such that $\sigmaa^*(a;0) = 0$, $\sigmaa^*(a;1) = 1$, and $\sigmad^*(a;p)= \mathbb{I}\{p>1/2\}$, is a pure strategy equilibrium.}
\label{fig:wishbone}
\vspace{-0.5em}
\end{figure}

\noindent This region, illustrated in Fig. \ref{fig:q10a_annotated}, is further subdivided into four subregions with the following interpretations: 
\begin{figure}[b]
\begin{center}
        \includegraphics[width=0.65\columnwidth]{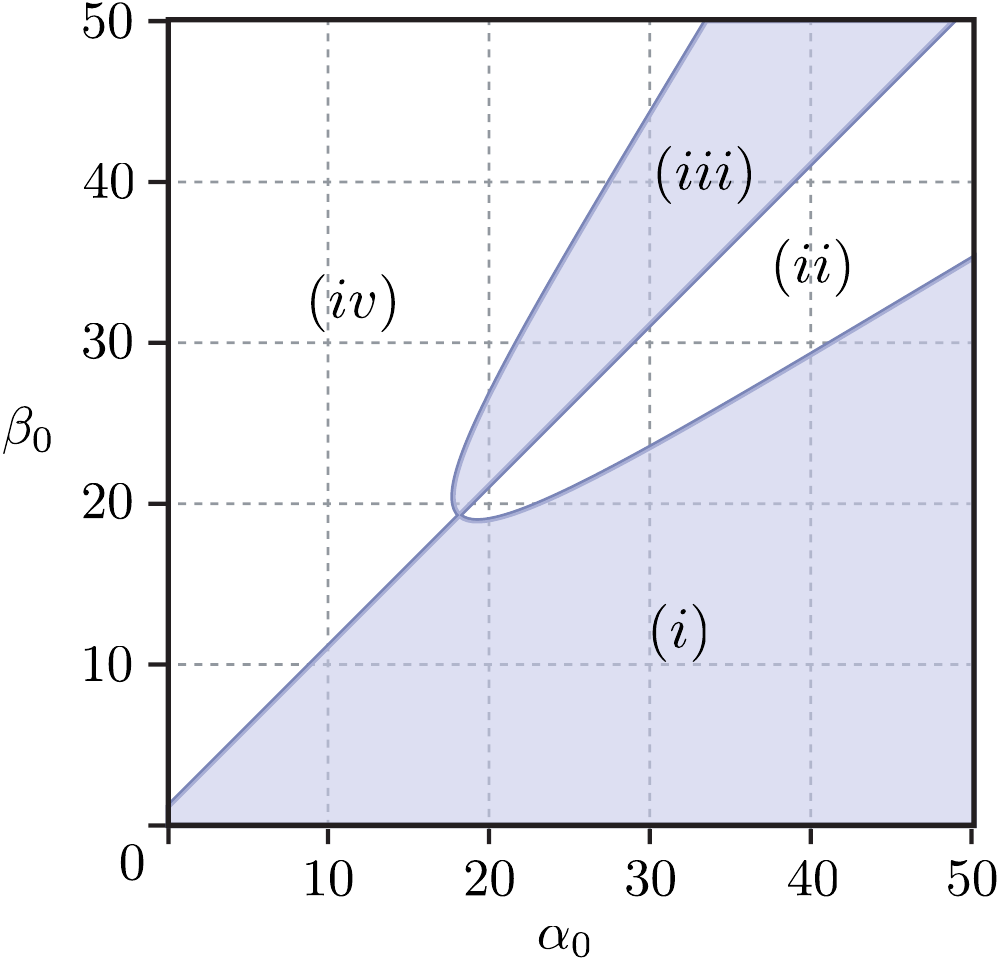}
\caption{Region for $(\alpha_0,\beta_0)$ such that $\sigmaa^*(a;\type=1) = 1$ when $\sigmad^*(a;p)= \mathbb{I}\{p>1/2\}$.}
\label{fig:q10a_annotated}
\end{center}
\end{figure}
\begin{enumerate}
\item[(i)] In this subregion, $\alpha_0$ is high relative to $\beta_0$ so target $a$ looks sufficiently desirable even if it gets caught. 
\item[(ii)] In this subregion, $\alpha_0$ is still sufficiently larger than $\beta_0$, so the uncertain reward appears better to the attacker than the certain reward. However, the attacker believes it to be very likely that $p > 1/2$ (due to a higher $\alpha_0+\beta_0$), and therefore would get caught should it choose $a$, outweighing its gain in reward. The capture cost $c$ deters the attacker from choosing $a$ at this point.
\item[(iii)] Analogous to case (ii), the attacker is sufficiently confident that the defender will defend target $b$, and thus the attacker chooses target $a$, even though the expected reward of the uncertain target is lower than the certain target.
\item[(iv)] Analogous to (i), the attacker is confident that $a$ will yield no reward, and thus it does not choose $a$.
\end{enumerate}
The reasoning for $\sigmaa^*(a;\type=0) = 0$ (derived from the upper bound condition in Lemma \ref{lem:equil}) follows identically. Combining the lower and upper bounds, the equilibrium holds in the intersection of the two regions, as illustrated by Fig. \ref{fig:wishbone}.

 \begin{figure*}[t]
    \centering
    \begin{subfigure}{0.24\textwidth}
        \includegraphics[width=0.9\linewidth]{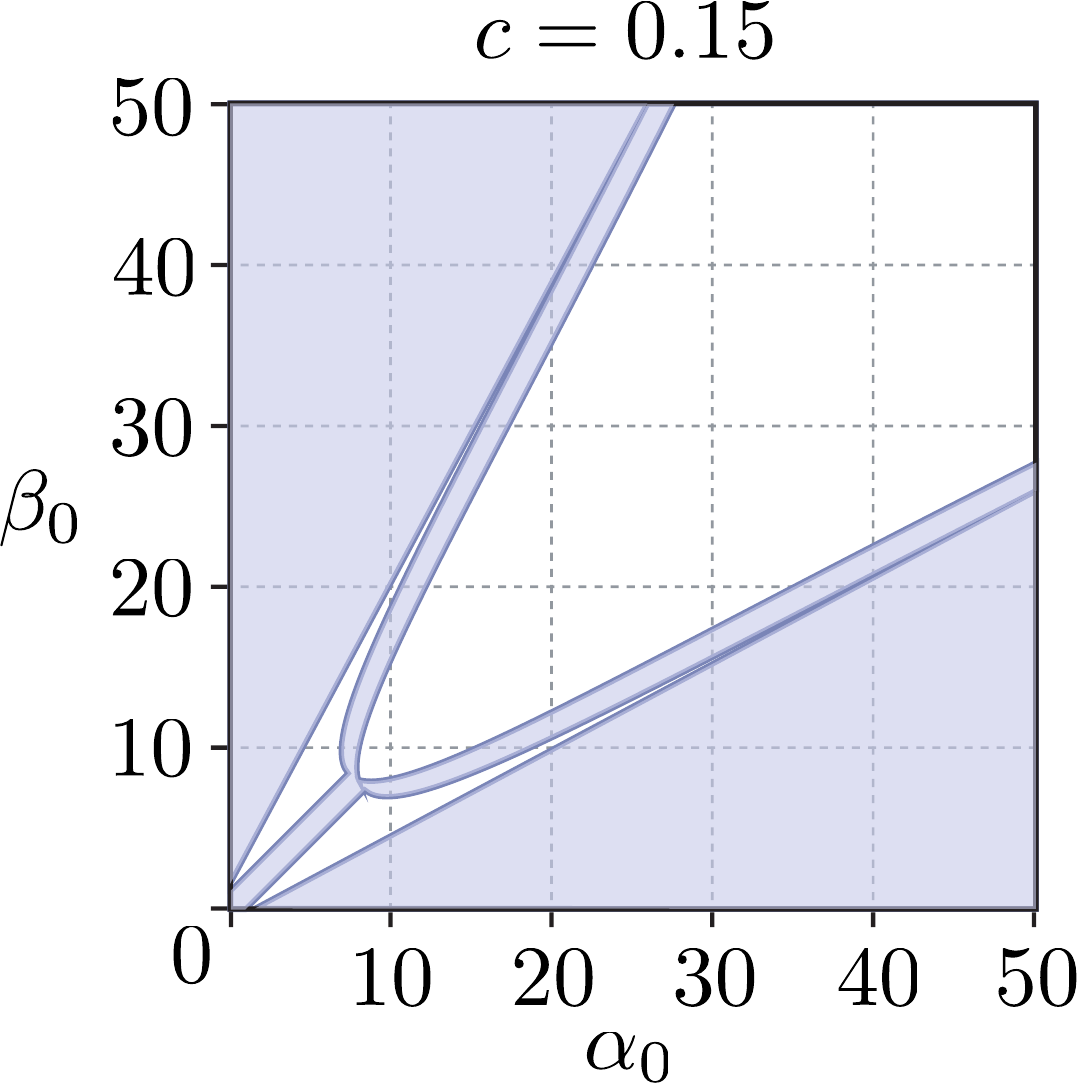}
        \caption{Regions under $c=0.15$.}
        \label{fig:c015}
    \end{subfigure}%
    \hspace{0.15\columnwidth}%
     \begin{subfigure}{0.24\textwidth}
    \centering
        \includegraphics[width=0.9\linewidth]{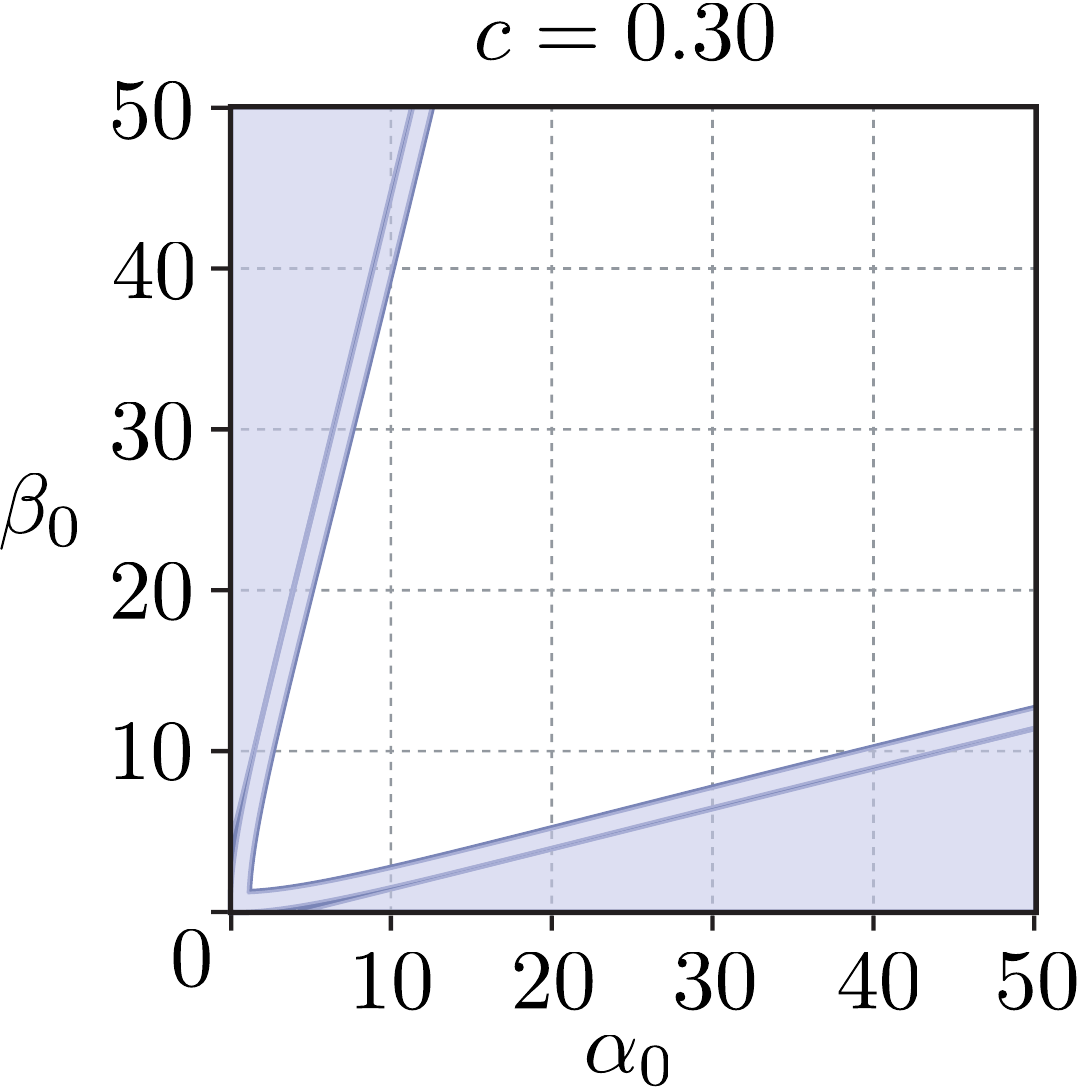}
        \caption{Regions under $c=0.30$.}
        \label{fig:c030}
    \end{subfigure}%
    \hspace{0.15\columnwidth}%
     \begin{subfigure}{0.24\textwidth}
    \centering
        \includegraphics[width=0.9\linewidth]{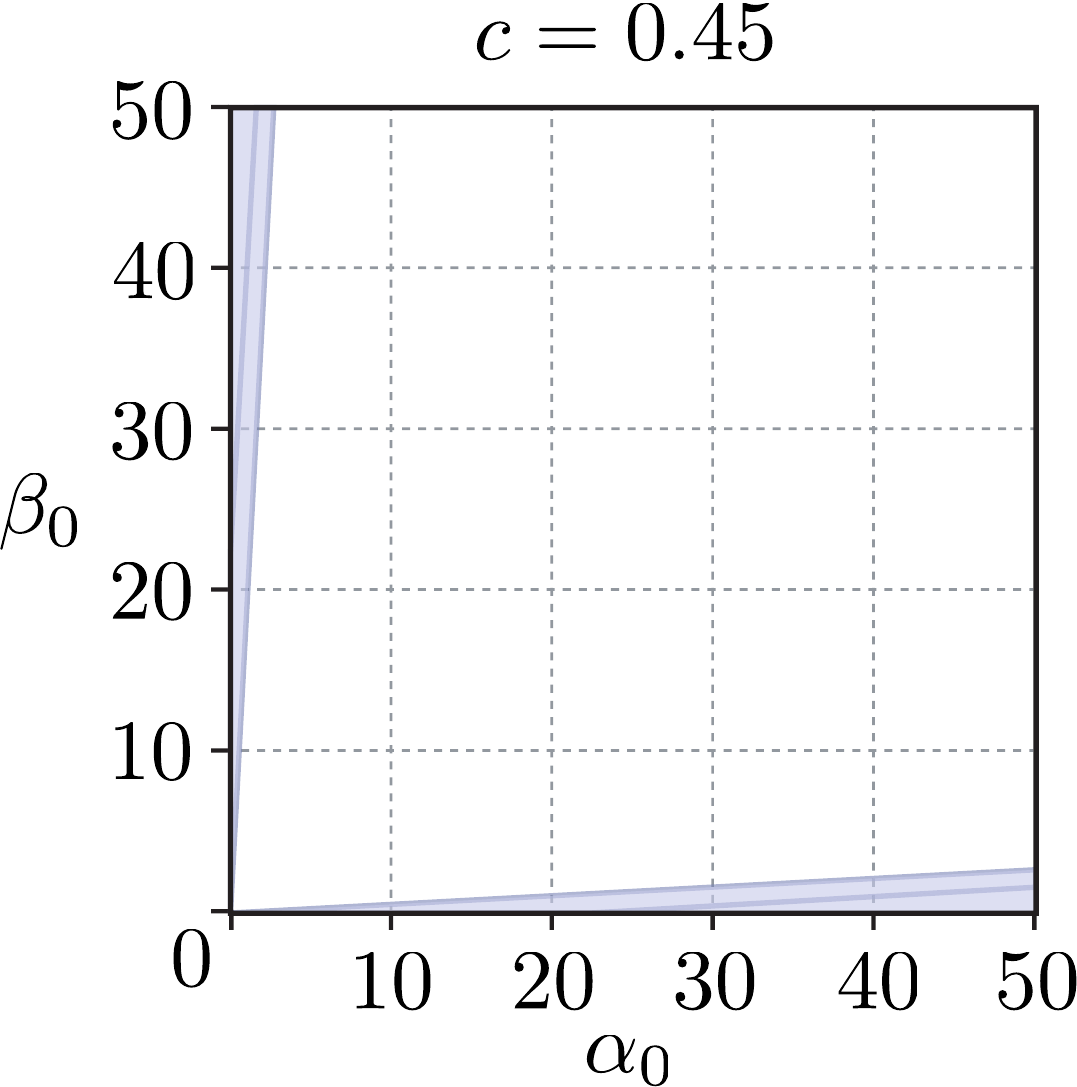}
        \caption{Regions under $c=0.45$.}
        \label{fig:c045}
    \end{subfigure}%
    \caption{Sensitivity analysis for the pure-strategy equilibrium regions of Lemma \ref{lem:equil} as a function of $c$ (for $R=1$, $\gamma = 1/2$).}
    \label{fig:varyc}
\end{figure*}

We now vary the capture cost to see how the equilibrium-supporting regions change. In particular, we look at the three regions in Lemma \ref{lem:equil} as we increase $c$. As $c$ is increased, see Figs. \ref{fig:c015} -- \ref{fig:c045}, the attacker is increasingly concerned with getting caught. This requires the attacker to have an increasing level of confidence that it will get $R$ from choosing $a$ in order to outweigh the higher risk of getting caught. 
%
For large enough $c$, such that $R-c < \gamma R$, all pure strategy equilibrium regions vanish. 
Committing to a choice of either $a$ or $b$ (a pure strategy) becomes too risky, causing the attacker to want to (either partially or fully) mix between targets. In other words, a sufficiently large $c$ will partially deter an attack on either of the targets. A full characterization of these mixed strategies is left for future work.

\section{CONCLUDING REMARKS}
\label{sec:conclusion}

Motivated by cyber security settings, we have introduced a simple asymmetric information game model for describing the influence of a learner's (the attacker) single private sample on the inference process of an observing agent (the defender). 
The subsequent game admits at most one pure strategy equilibrium which, depending on the parameters of the game, takes different forms. We illustrated that, even in the case where the attacker is confident that target $a$ will produce a reward higher than the known reward associated with the other target, it is not necessarily optimal for the attacker to attack target $a$ as it has an increased chance of being caught. 
Future work includes considering multiple unknown targets and allowing the attacker to make sampling decisions. If these choices (but not the received samples) are observable to the defender, then the game becomes a signaling game.

\bibliographystyle{IEEEtran}  
\bibliography{references}

\appendices

\vspace{-0.5em}
\section{Proofs}
\subsection{Proof of Lemma \ref{lem:eqconds}}
\label{app:lem:eqconds}

Denote by $\ua(a,a)$ the attacker's subjective payoff when both players select $a$ as dictated by the subjective payoffs in Table \ref{tab:sub} (a), similarly for $\ua(a,b)$, $\ua(b,a)$, and $\ua(b,b)$. Also, denoting $\Ex_{\pa}\big[\sigmad^*(a;\pa)\mid \type\big]$ by $\Ex[\sigmad^*]$ and substituting in subjective payoffs, the attacker's expected payoff is
\begin{align*}
	&\Ex_{p}\big[\ua(\sigmaa ,\sigmad ^*)\mid\type\big] \\
	&=\sigmaa \bigg(\ua(a,a)\Ex[\sigmad ^*] +\ua(a,b)\big(1-\Ex[\sigmad ^*]\big)\bigg)\\
	&\hspace{2em}+(1-\sigmaa )\bigg(\ua(b,a)\Ex[\sigmad ^*] + \ua(b,b)\big(1-\Ex[\sigmad ^*]\big)\bigg)\\
	&=\sigmaa \bigg(\bigg(\frac{\alpha_0+\type}{\alpha_0+\beta_0+1}R-c\bigg)\Ex[\sigmad ^*]\\
	&\hspace{5em} +\frac{\alpha_0+\type}{\alpha_0+\beta_0+1}R\big(1-\Ex[\sigmad ^*]\big)\bigg)\\
	&\hspace{2em}+(1-\sigmaa )\big(\gamma R\Ex[\sigmad ^*] + (\gamma R-c)\big(1-\Ex[\sigmad ^*]\big)\big)\\
	&= \gamma R\Ex[\sigmad ^*] + (\gamma R-c)(1-\Ex[\sigmad ^*]) \\
	&\hspace{2em}+ \sigmaa \bigg(\bigg(\frac{\alpha_0+\type}{\alpha_0+\beta_0+1} - \gamma\bigg)R+c\big(1-2\Ex[\sigmad ^*]\big)\bigg)
\end{align*}
The first two terms do not influence the $\argmax$ and thus one can equivalently maximize $\sigmaa \big(\big(\frac{\alpha_0+\type}{\alpha_0+\beta_0+1} - \gamma\big)R+c\big(1-2\Ex[\sigmad ^*]\big)\big)$. 
Similarly, using Table \ref{tab:sub} (b) and denoting $\Ex_{\type}\big[\sigmaa ^*(\type)\mid \pa\big]$ by $\Ex[\sigmaa ^*]$, the defender's expected cost is
\begin{align*}
	\Ex_{\type}\big[\ud(\sigmaa ^*,\sigmad )\mid \pa\big]\\
	&\hspace{-7em}=\sigmad \big(\ud(a,a)\Ex[\sigmaa ^*] + \ud(b,a)(1-\Ex[\sigmaa ^*])\big)\\ 
	&\hspace{-5em}+ (1-\sigmad )\big(\ud(a,b)\Ex[\sigmaa ^*] + \ud(b,b)(1-\Ex[\sigmaa ^*])\big)\\
	&\hspace{-7em}=\sigmad \big((\pa R-c)\Ex[\sigmaa ^*] + \gamma R(1-\Ex[\sigmaa ^*])\big)\\
	&\hspace{-5em}+ (1-\sigmad )\big(\pa R\Ex[\sigmaa ^*] + (\gamma R-c)(1-\Ex[\sigmaa ^*])\big)\\
	&\hspace{-7em}= \gamma R-c+\Ex[\sigmaa ^*](\pa R - \gamma R+c)+c\sigmad \big(1-2\Ex[\sigmaa ^*]\big)
\end{align*}
Again, the constants do not influence the $\argmin$ and one can equivalently minimize $\sigmad \big(1-2\Ex[\sigmaa ^*]\big)$.\QED

\subsection{Proof of Lemma \ref{lem:equil}}
\label{app:lem:equil}

Throughout the proof, let $q_{\type}=\sigmaa(a;\type)$. The attacker's expected strategy from the defender's perspective is $\Ex_{\type}\big[q_\type\mid \pa\big] = (1-\pa)q_0 + \pa q_1$. There are four possible cases:

\vspace{0.25em}
\noindent\textbf{1.i)} \emph{$q_0 = q_1 = 1$}: Assume the attacker plays $a$ regardless of its type, \emph{i.e.}, $q_\type = 1$ $\forall$ $\type$. Thus $E_\type[q_\type\mid p] = 1$ and by Lemma \ref{lem:eqconds}, the defender's best response is $\sigmad^*(a;p) = 1$ $\forall$ $p$. The attacker does not deviate iff
\begin{align*}
	\bigg(\frac{\alpha_0+\type}{\alpha_0+\beta_0+1} - \gamma \bigg)R+c\big(1-2\Ex_{\pa}\big[\sigmad^*(\pa)\mid \type\big]\ge0\\
	\Leftrightarrow \bigg(\frac{\alpha_0+\type}{\alpha_0+\beta_0+1} - \gamma\bigg)R \ge c
\end{align*}

\vspace{0.25em}
\noindent\textbf{1.ii)} \emph{$q_0 = q_1 = 0$}: Assume the attacker plays $b$ regardless of its type. Then $E_\type[q_\type\mid p] = 0$ and the best response of the defender is $\sigmad^*(a;p) = 0$ $\forall$ $p$. Similar to case (1.i), the attacker does not deviate iff $\big(\gamma - \frac{\alpha_0+\type}{\alpha_0+\beta_0+1}\big)R \ge c$.

\vspace{0.25em}
\noindent\textbf{1.iii)} \emph{$q_0 = 1, q_1 = 0$}: Assume the attacker plays $q_0 = 1$, $q_1 = 0$. Then $E_\type[q_\type\mid p]  = 1-\pa $. The best response (almost surely) for the defender is $\sigmad^*(a;p)= \mathbb{I}\{p<1/2\}$. By Lemma \ref{lem:eqconds}, the attacker plays $q_0 = 1$ iff
\begin{align*}
	\Ex_{\pa}\big[\sigmad^*(a;\pa)\mid 0\big] < \frac{1}{2}\bigg(1+\frac{R}{c}\Big(\frac{\alpha_0}{\alpha_0+\beta_0+1} - \gamma\Big)\bigg)
\end{align*}
or equivalently, using the fact that $\Ex_{\pa}\big[\sigmad^*(a;\pa)\mid \type=0\big] = \Ex[\mathbb{I}\{\pa <1/2\}\mid \type=0] = P(\pa <1/2\mid \type=0)$,
\begin{align}
	I_{1/2}(\alpha_0,\beta_0+1) \le \frac{1}{2}\bigg(1+\frac{R}{c}\Big(\frac{\alpha_0}{\alpha_0+\beta_0+1} - \gamma\Big)\bigg)\label{eq:case1_2_1}.
\end{align}
By a similar argument, the attacker plays $q_1 = 0$ iff
\begin{align}
	\nonumber I_{1/2}(\alpha_0+1,\beta_0) &\ge \frac{1}{2}\bigg(1+\frac{R}{c}\Big(\frac{\alpha_0+1}{\alpha_0+\beta_0+1} - \gamma\Big)\bigg)\\
	&\hspace{-8em}= \frac{1}{2}\bigg(1+\frac{R}{c}\Big(\frac{\alpha_0}{\alpha_0+\beta_0+1} - \gamma\Big)\bigg) + \frac{R}{2c(\alpha_0+\beta_0+1)}\label{eq:case1_2_2}
\end{align}
Conditions (\ref{eq:case1_2_1}) and (\ref{eq:case1_2_2}) imply that,
\begin{align}
	\nonumber I_{1/2}(\alpha_0+1,\beta_0) - I_{1/2}(\alpha_0,\beta_0+1)\ge \frac{R}{2c(\alpha_0+\beta_0+1)}
\end{align}
Using consecutive neighbor identities of the regularized incomplete beta function, 
the above inequality becomes
\begin{align*}
	-\frac{1}{B(\alpha_0+1,\beta_0+1)}\Big(\frac{1}{2}\Big)^{\alpha_0+\beta_0-1}\ge \frac{R}{c}
\end{align*}
which can never be satisfied, thus at least one of Eqs. (\ref{eq:case1_2_1}) and (\ref{eq:case1_2_2}) is violated.

\vspace{0.25em}
\noindent\textbf{1.iv)} \emph{$q_0 = 0, q_1 = 1$}: Assume the attacker plays $q_0 = 0$, $q_1 = 1$. Then $E_\type[q_\type\mid p]  = \pa $. The defender's best response is $\sigmad^*(a;p)= \mathbb{I}\{p>1/2\}$. The attacker plays $q_0 = 0$ iff
\begin{align*}
	\Ex_{\pa}\big[\sigmad^*(a;\pa)\mid 0\big] > \frac{1}{2}\bigg(1+\frac{R}{c}\Big(\frac{\alpha_0}{\alpha_0+\beta_0+1} - \gamma\Big)\bigg)
\end{align*}
Since $\Ex[\sigmad^*(a;p)\mid \type=0] = 1-P(\pa \le 1/2\mid \type=0)$, an equivalent condition is
\begin{align*}
	I_{1/2}(\alpha_0,\beta_0+1) \le \frac{1}{2}\bigg(1+\frac{R}{c}\Big(\gamma-\frac{\alpha_0}{\alpha_0+\beta_0 + 1} \Big)\bigg).
\end{align*}
Similarly, the attacker plays $q_1 = 1$ iff
\begin{align*}
	I_{1/2}(\alpha_0+1,\beta_0) \ge \frac{1}{2}\bigg(1+\frac{R}{c}\Big(\gamma-\frac{\alpha_0+1}{\alpha_0+\beta_0+1}\Big)\bigg).
\end{align*}
Or equivalently, by consecutive neighbor identities, 
\begin{align*}
	I_{1/2}(\alpha_0,\beta_0) \in\\
	&\hspace{-6em}\Bigg[\frac{1}{2}\bigg(1+\frac{R}{c}\Big(\gamma-\frac{\alpha_0+1}{\alpha_0+\beta_0+1}\Big)\bigg) + \frac{\Big(\frac{1}{2}\Big)^{\alpha_0+\beta_0}}{\alpha_0B(\alpha_0,\beta_0)},\\ 
	&\hspace{-5.4em}\frac{1}{2}\bigg(1+\frac{R}{c}\Big(\gamma-\frac{\alpha_0}{\alpha_0+\beta_0 + 1} \Big)\bigg) - \frac{\Big(\frac{1}{2}\Big)^{\alpha_0+\beta_0}}{\beta_0B(\alpha_0,\beta_0)}\Bigg]
\end{align*}
which can be seen to be non-empty by selecting $R$ and $c$ appropriately.  \QED

\end{document}